\documentclass[conference]{IEEEtran}
\IEEEoverridecommandlockouts
\usepackage[utf8]{inputenc}
\usepackage{cuted}
\usepackage{subcaption}
\usepackage{cite}
\usepackage{amsmath,amssymb,amsfonts}
\usepackage{algorithmic}
\usepackage{graphicx}
\usepackage{textcomp}
\usepackage{xcolor}
\usepackage[export]{adjustbox}
\usepackage{xurl}
\usepackage{comment}
\usepackage{paralist} 
\usepackage{enumitem} 
\usepackage[bookmarks=false]{hyperref}
\usepackage[T1]{fontenc}

\usepackage{tabularx}
\usepackage{multirow}
\usepackage{makecell}
\usepackage{mathtools}

\title{Spatiotemporal Impact Assessment of Hurricanes on Electric Power Systems
\thanks{This work is supported by NSF Career Award 1944142.}
}

\author{\IEEEauthorblockN{Abodh Poudyal$^{\dagger}$, Anamika Dubey}
\IEEEauthorblockA{Washington State University \\ Pullman, Washington, USA\\
Email: $^\dagger$abodh.poudyal@wsu.edu}

\and
\IEEEauthorblockN{Vishnu Iyengar}
\IEEEauthorblockA{Lakeside School \\ Seattle, Washington, USA}

\and
 
\IEEEauthorblockN{Diego Garcia-Camargo}
\IEEEauthorblockA{Stanford University \\ Stanford, California, USA}
 }

\begin{document}
\bstctlcite{IEEEexample:BSTcontrol}
\maketitle
\thispagestyle{plain}
\pagestyle{plain}

\begin{abstract}
Severe windstorms such as hurricanes are the primary cause of extensive power grid damages resulting in widespread customer outages and expensive recovery. 
This paper aims to develop a probabilistic impact assessment framework to model and quantify the spatiotemporal impacts of windstorms such as hurricanes on the bulk power grid. The variations in hurricane trajectory and wind speed are modeled using historical data from past events in the US.  The impacts of individual power grid components are modeled using fragility curves typically obtained using historical outage data. Finally, the system losses are modeled using a loss metric quantifying the total load disconnected due to the impact of the hurricane as it travels inland. The simulation is performed on the 2000-bus synthetic Texas grid mapped on the geographical footprint of Texas. The simulation results show that the loss increases significantly for a few time steps when the hurricane's wind field is intense and saturates gradually when the hurricane's intensity decays while traversing further inland.
\end{abstract}


\section{Introduction}\label{intro}
Hurricanes, one of the most severe conditions of \textit{tropical storms} with wind-speed exceeding 75 miles per hour (mph), are reported to be the primary cause of power outages due to extreme weather events resulting in an economic loss of over \$1.5 trillion in the US alone~\cite{cost_NOAA}. For example, Hurricane Ida is estimated to have cost \$16 to 24 billion in flooding damage in the Northeastern US. About 1.2 million customers were left without power -- it took almost 15 days to restore the electric power entirely. These events are categorized as high impact low probability (HILP) events or black swan events. However, the frequency of the events carrying the highest impacts has drastically increased, costing around \$22 billion in climate-related disasters in 2020 alone in the USA~\cite{OCM}. Thus, with the growing intensity and frequency of extreme weather events, there is a critical need to characterize and quantify the impacts of these HILP events on the electric power grid. 


Although long-range weather prediction models have improved in recent years, they have not been adequately utilized to evaluate climate impacts on the power grid. For example, there have been significant efforts to forecast the hurricane's tracks, where the forecast error for the tracking has reduced from 300 nautical miles (nm) in 1990 to about 100 nm in 2016~\cite{2018}. A weather-grid impact model provides a much-needed capability to the grid planners in adapting the existing grid to become a resilient grid ~\cite{kwasinski2019hurricane,zhang2019spatial}.  Without a weather-grid impact model, planning for long-term grid resilience may not be well informed or effective.  Thus, there is a critical need to develop an appropriate weather-impact model for the power grid by adequately modeling the associated spatio-temporal properties of the extreme weather event.

A large body of existing work studies the effects of hurricanes on the power grid, as reviewed below. A majority of the existing work considers hurricanes as a function of the wind-speed; however, other dynamics associated with the hurricanes that are crucial to analyze their impacts on the power grid are ignored~\cite{panteli2015modeling, 8848458}. A grid resilience model considering hurricane-induced damages and integrated renewable resources is presented in~\cite{watson2019modeling}. However, the approach is not generic as it considers a single case study on Hurricane Harvey. In~\cite{eskandarpour2016machine}, a data-driven outage prediction method is proposed, which uses a decision boundary method to identify the operational state of grid components. However, the authors do not report any analysis to quantify the damages or impacts on the grid. Authors in~\cite{nguyen2019assessing} present a framework to analyze the impacts of energy storage towards improving the resilience of distribution systems against hurricanes. However, since the hurricane's radius is greater than an entire distribution grid at any instant time, such analysis would not be meaningful for a distribution grid alone. Finally, ~\cite{reilly2017hurricanes} introduces a community and customer-based perspective in improving the resilience of the power grid against hurricanes; however, the spatiotemporal effects of hurricanes are not modeled.

The existing literature lacks in adequately modeling the spatio-temporal impacts of dynamic hurricane events and their impacts on the large power grid. The impact assessment model should capture different uncertainties associated with the stochastic nature of the event and their impacts on the individual power grid components and the power system as a whole. With these considerations, this paper aims to develop a probabilistic weather-impact assessment framework that appropriately models the stochastic nature of hurricanes and their time-varying impacts on the large-scale power grid. First, we present an approach to generate dynamic hurricane scenarios considering different hurricane tracks that are mapped onto a transmission grid with geospatial information. The hurricane parameters are sampled from actual hurricanes that have previously occurred in the USA. Next, we develop a probabilistic weather-impact assessment model using Monte-Carlo simulations considering a large number of hurricanes with different tracks and at different time steps. The damage levels are mapped with the respective fragility curves of the transmission lines to visualize the spatiotemporal effects of the event. Finally, a spatiotemporal loss metric is proposed to quantify the impacts of probabilistic hurricane scenarios on the power grid.




\vspace{-5pt}

\section {Modeling}\label{sec:hurricane_power_model}

This section details the modeling of hurricane dynamics, its impacts on the power grid components, and the approach to model the spatiotemporal effects on the power grid.
\subsection{Hurricane Wind Field Model}
\begin{figure}[ht]
    \centering
    \includegraphics[width=0.8\linewidth]{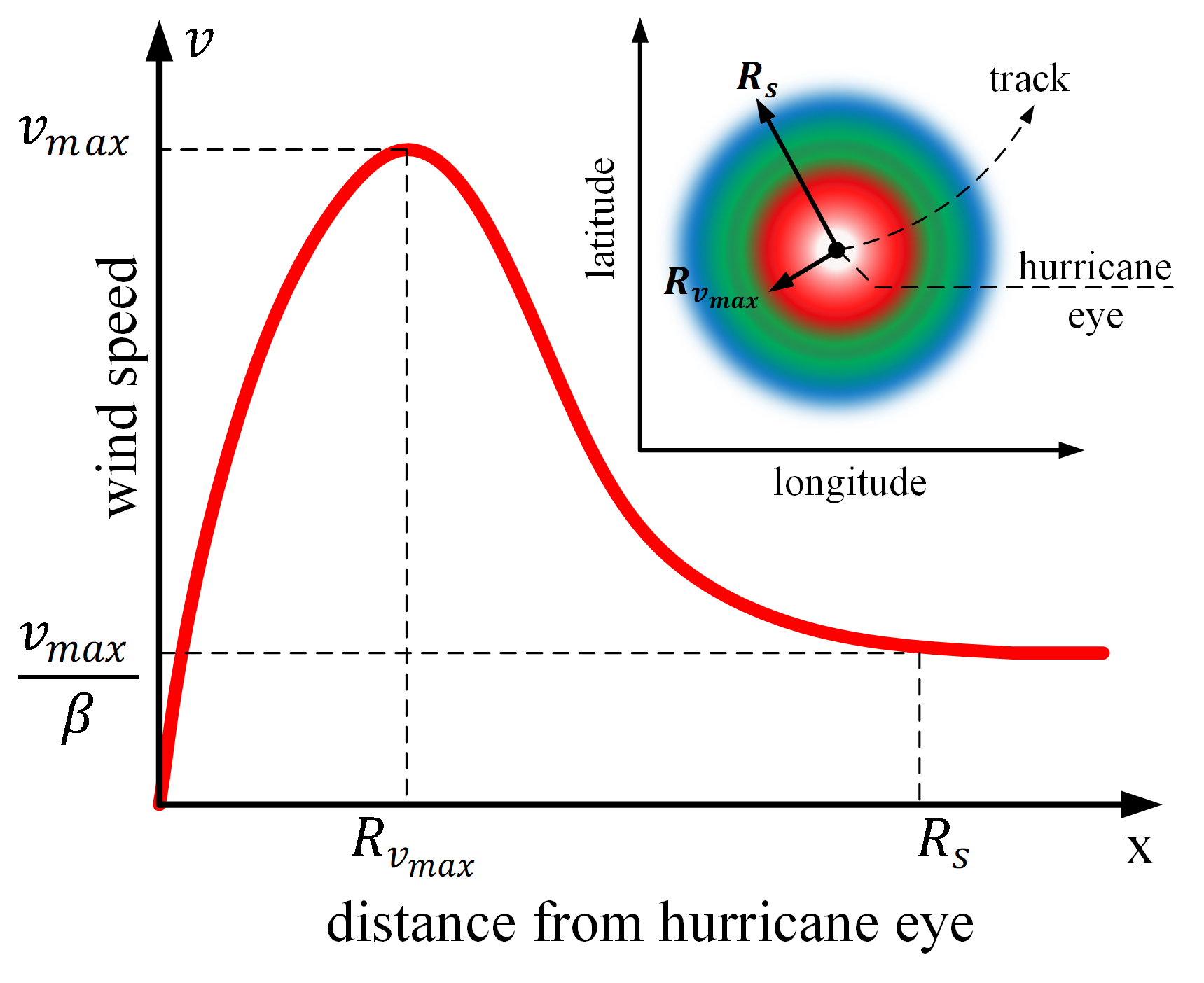}
    \caption{Static gradient wind field of a typical hurricane.}
    \label{fig:static}
    \vspace{-5pt}
\end{figure}

The damage created by the hurricane is due to high-intensity wind speed, spread around the region of its eye, that changes along its path as it moves inland from the location of its landfall~\cite{hurricane_report}. With hurricane eye as the reference, the static gradient wind field model of the hurricane is defined using three variables --- maximum sustained wind speed of the hurricane ($v_{max}$) in knots, the distance to $v_{max}$ from the hurricane eye ($R_{v_{max}}$) in nautical miles (nmi), and the radius of the hurricane from the hurricane eye ($R_s$) in nmi~\cite{2012}. The static model represents the wind field model of a hurricane at a particular instance of time as shown in Fig.~\ref{fig:static}. A piecewise mathematical function of the gradient wind field model shown in Fig.~\ref{fig:static} can be represented using (1)~\cite{Javanbakht2014}.
\vspace{-10pt}

\begin{equation} \label{eq:static_eq}
\small
\begin{aligned}
&v(x)= \begin{dcases} K \times v_{max}(1-exp{[-\Psi x]}) & 0 \leq x<R_{v_{max}} \\
v_{max} \exp \left[- \Lambda \left(x-R_{v_{max}}\right)\right] & R_{v_{max}} \leq x \leq R_{s} \\
0 & x>R_{s}\end{dcases} \\
&\Psi=\frac{1}{R_{v_{max}}} \ln \left(\frac{K}{K-1}\right), K>1;
\quad \Lambda = \frac{\ln \beta}{R_{s}-R_{v_{max}}}
\end{aligned}
\end{equation}

\noindent
where, $x$ is the distance from the hurricane eye , $K$ is a constant depending on the nature of the hurricane, $\beta$ is the factor by which the maximum sustained wind speed decreases at the boundary of the hurricane, $R_s$. It is also assumed that the hurricane has no effect outside of its boundary. 


The dynamic behavior of a hurricane can be incorporated by simulating the static model at several discrete time intervals. Having this dynamic wind field allows us to track the severity of a hurricane as it moves along its path after landfall. The hurricane has minimum wind speed and maximum pressure at the location of the eye given by~\cite{2012}:
\vspace{-5pt}
\begin{equation}
    \Delta P_{h}^{0,\zeta} = \sqrt{\frac{2.636+0.0394899\phi^{0,\zeta} - ln(R_{v_{max_h}}^{0,\zeta})}{5.086\times10^{-4}}}
    \label{eq:landfall_pressure}
\end{equation}

\noindent
where, $\phi^{0,\zeta}$ is the latitude of the location of landfall for track $\zeta$, and $\Delta P_{h}^{0,\zeta}$ is the pressure of the hurricane eye at time step $t=0$ for hurricane $h$ and track $\zeta$. Here, it is assumed that the landfall always occurs at time step $t=0$ for each $\zeta$. Thus, for each hurricane $h$, the pressure at the eye location for each time step $t$ and track $\zeta$ can then be calculated as:

\begin{equation}
    \Delta P_{h}^{t,\zeta} = \Delta P_{h}^{0,\zeta} e^{-\alpha t}
    \label{eq:time_pressure}
\end{equation}

\noindent
Here, (\ref{eq:time_pressure}) represents the hurricane decay model with land decay parameter $\alpha$. Using the above equations and for each $t$, new values of $\{R_{v_{max}}^{t,\zeta}$, $R_{s}^{t,\zeta}$, $v_{max}^{t,\zeta}\}_h$ are obtained for each hurricane scenario, $h$, based on their values in the previous time step and the known coordinates of the forecasted hurricane tracks obtained for each $\zeta$. Using these new values, a static wind field is generated for each $t$ and $\zeta$, resulting in a dynamic wind field model~\cite{Javanbakht2014}.

\subsection{Power Systems Impact Model}
The hurricane model described above will have a spatiotemporal effect on the power grid. That is, each of the spatially distributed power grid components will experience different wind speeds at time instances. To capture spatiotemporal effects, we modify the hurricane impact model presented in~\cite{Javanbakht2014} by introducing additional time and track variables. Accordingly, the wind speed experienced by the transmission line $l$, at discrete time-interval $t$ and track $\zeta$ is given by (4).
\vspace{-5pt}

\small
\begin{equation}
\begin{aligned}
&\Gamma_{l, h}^{t,\zeta}= \begin{cases}v_{max_{h}}^{t,\zeta} & {d_{\min}^{t,\zeta}(l, h) \leq R_{v_{max_h}}^{t,\zeta} \leq d_{\max }^{t,\zeta}(l, h)} \\
\max\left(\begin{array}{c} v^{t,\zeta} \left(\gamma_1\right) \\
v^{t,\zeta}\left(\gamma_2 \right)
\end{array}\right) & \text {otherwise} \end{cases}\\
& \gamma_1 = d_{\min }^{t,\zeta}(l, h) \mid \{R_s^{t,\zeta}, R_{v_{max}}^{t,\zeta}, v_{max}^{t,\zeta}\}_h\\
& \gamma_2 = d_{\max }^{t,\zeta}(l, h) \mid \{R_s^{t,\zeta}, R_{v_{max}}^{t,\zeta}, v_{max}^{t,\zeta}\}_h\\
\end{aligned}
\label{eq:HIM}
\end{equation}

\normalsize
\noindent
where, $\Gamma_{l, h}^{t,\zeta}$ is the maximum wind speed experienced by line $l$ for the hurricane scenario, $h$, in track $\zeta$ and at time step $t$. $d_{min}^{t,\zeta}(.)$ and $d_{max}^{t,\zeta}(.)$ represent the minimum and maximum distance of $l$ from the eye of the hurricane $h$ for track $\zeta$ at time step $t$. $v^{t,\zeta}(.)$ is calculated using (\ref{eq:static_eq}). For simplicity, the lines connected between any two buses are assumed to be straight lines and the wind speed across the entire section of a single transmission line is assumed to be the same.  

It is to be noted that only a few of the lines, $l$, will have severe impact depending on the intensity of the wind it experiences at that particular time instant. The wind speed experienced by each transmission line, $l$, can be mapped with its outage probability as suggested in~\cite{7801854}. It is to be noted that the wind speed, obtained from Eq.~\ref{eq:HIM}, varies in spatiotemporal dimension and hence, each $l$ will have different damage/outage probability at different time steps, $t$. Thus, the outage probability for a transmission line, $l$, experiencing a hurricane, $h$, at time $t$ is calculated using (5).

\vspace{-10pt}

\begin{equation} 
\begin{aligned}
&\mathbb{P}_{out}^{t,\zeta}(l,h)= \begin{dcases} 0 & \Gamma_{l, h}^{t,\zeta} < v_{cri} \\
\frac{\Gamma_{l, h}^{t,\zeta} - v_{cri}}{v_{col} - v_{cri}}  & v_{cri} \leq \Gamma_{l, h}^{t,\zeta} < v_{col} \\
1 & \Gamma_{l, h}^{t,\zeta} \geq v_{col} \end{dcases} \\
\end{aligned}
\label{eq:outage}
\end{equation}

\noindent
where, $v_{col}$ = 106.91 knots
is the wind speed at which the transmission line collapses, and $v_{cri}$ = 48.59 knots is the critical wind speed beyond which the transmission line is affected by the hurricane. Let $\delta_{l,h}^{t,\zeta}$ be the line status variable that is 1 if the line experiences an outage and 0 otherwise. If a transmission line $l'$ experiences an outage at time-step $t-1$ for any hurricane $h'$ and given track $\zeta'$, i.e., $\delta_{l',h'}^{t-1,\zeta'} = 1$  then $\mathbb{P}_{out}^{t',\zeta'}(l',h')$ = 1 i.e., $\delta_{l',h'}^{t',\zeta'} = 1$, $\forall t' > t - 1$. This ensures the time-varying impacts of the hurricane are modeled as it moves inland from the location of its landfall. 


\section{Methodology}\label{sec:method}
\begin{figure}[!t!]
    \centering
    \includegraphics[width=0.9\linewidth]{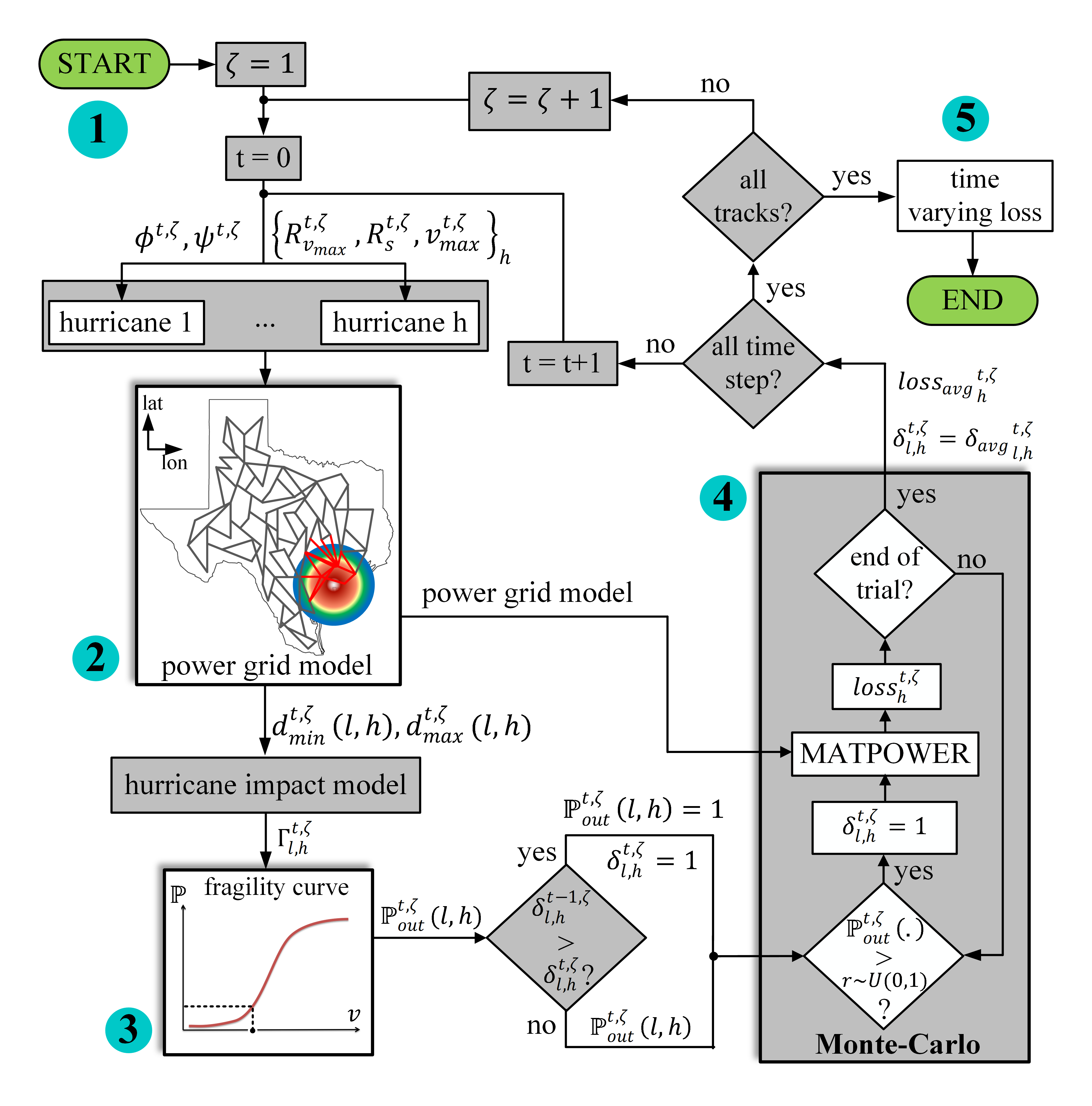}
    \caption{Overall flow diagram of generating spatiotemporal hurricane and assessing its impact on a power grid.}
    \label{fig:overall_picture}
    \vspace{-10pt}
\end{figure}

In this section, we describe the overall framework to assess the spatiotemporal impact of the hurricanes on the power grid as described in Fig.~\ref{fig:overall_picture}. First, we generate several hurricane scenarios with a known potential landfall location defined by the latitude ($\phi$) and longitude ($\psi$). The hurricane impact model defined in (\ref{eq:HIM}) is used to calculate the wind speed experienced by each transmission line. The outage probability of a line is then calculated using the fragility curves defined in (\ref{eq:outage}). Based on the damage probability distribution, Monte-Carlo simulations (MCS) are performed to determine the probabilistic system losses at the landfall location. The hurricane dynamics are used to simulate its movement inland. The component and system-level power grid damages are computed at several discrete time intervals. For each time step, the system loss metric is averaged over multiple hurricane scenarios and possible tracks. The details are provided in the following subsections.  

\subsection{Generating Hurricane Scenarios}
\begin{figure}
    \centering
    \begin{subfigure}[t]{0.49\linewidth}
        \centering
        \includegraphics[clip, trim=0cm 0.7cm 0.1cm 0.5cm, width=\linewidth]{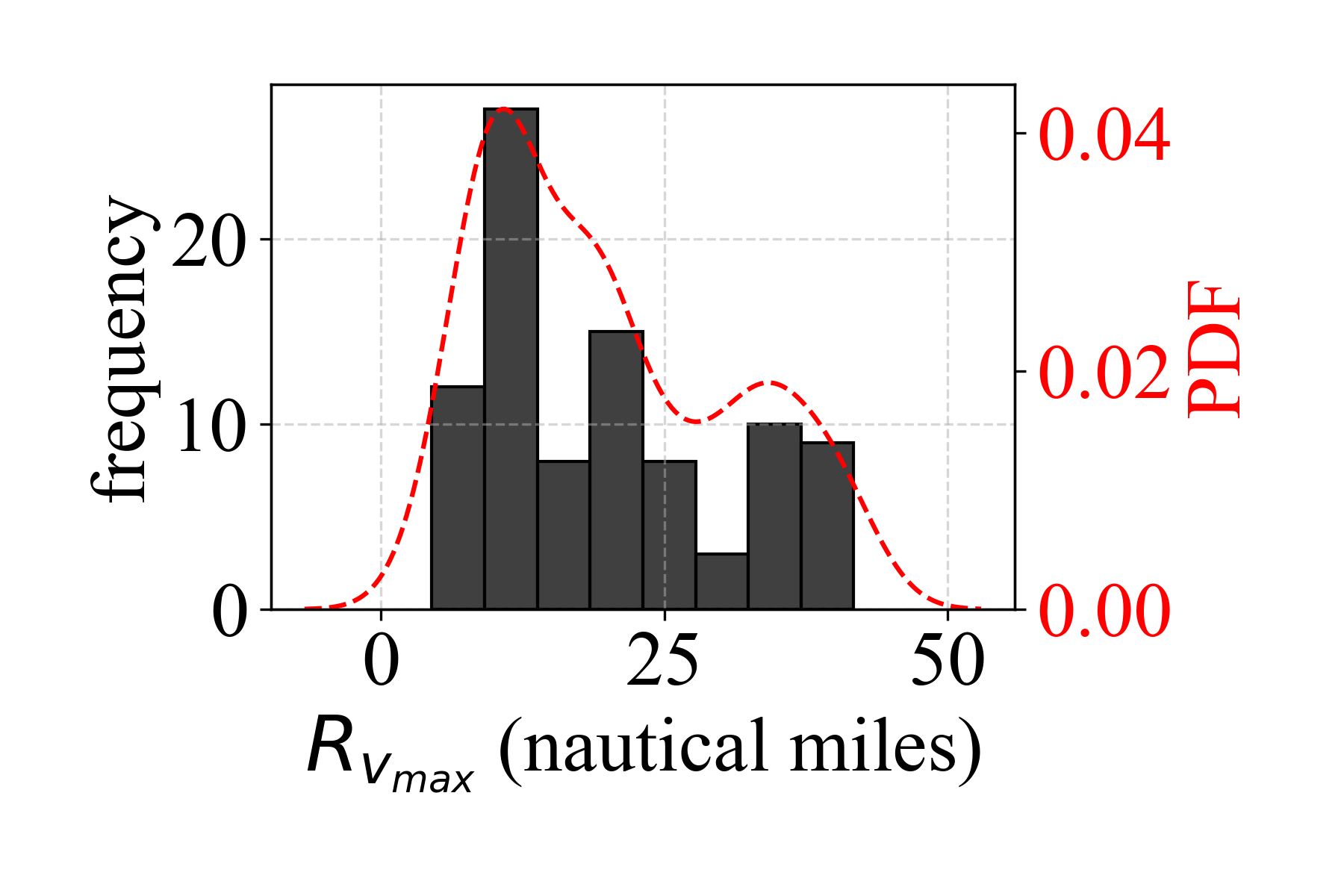}
        \vspace{-20pt}
        \caption{}
    \end{subfigure}%
    \hspace*{\fill}
    \begin{subfigure}[t]{0.49\linewidth}
        \centering
        \includegraphics[clip, trim=0cm 0.7cm 0.1cm 0.5cm,width=\linewidth]{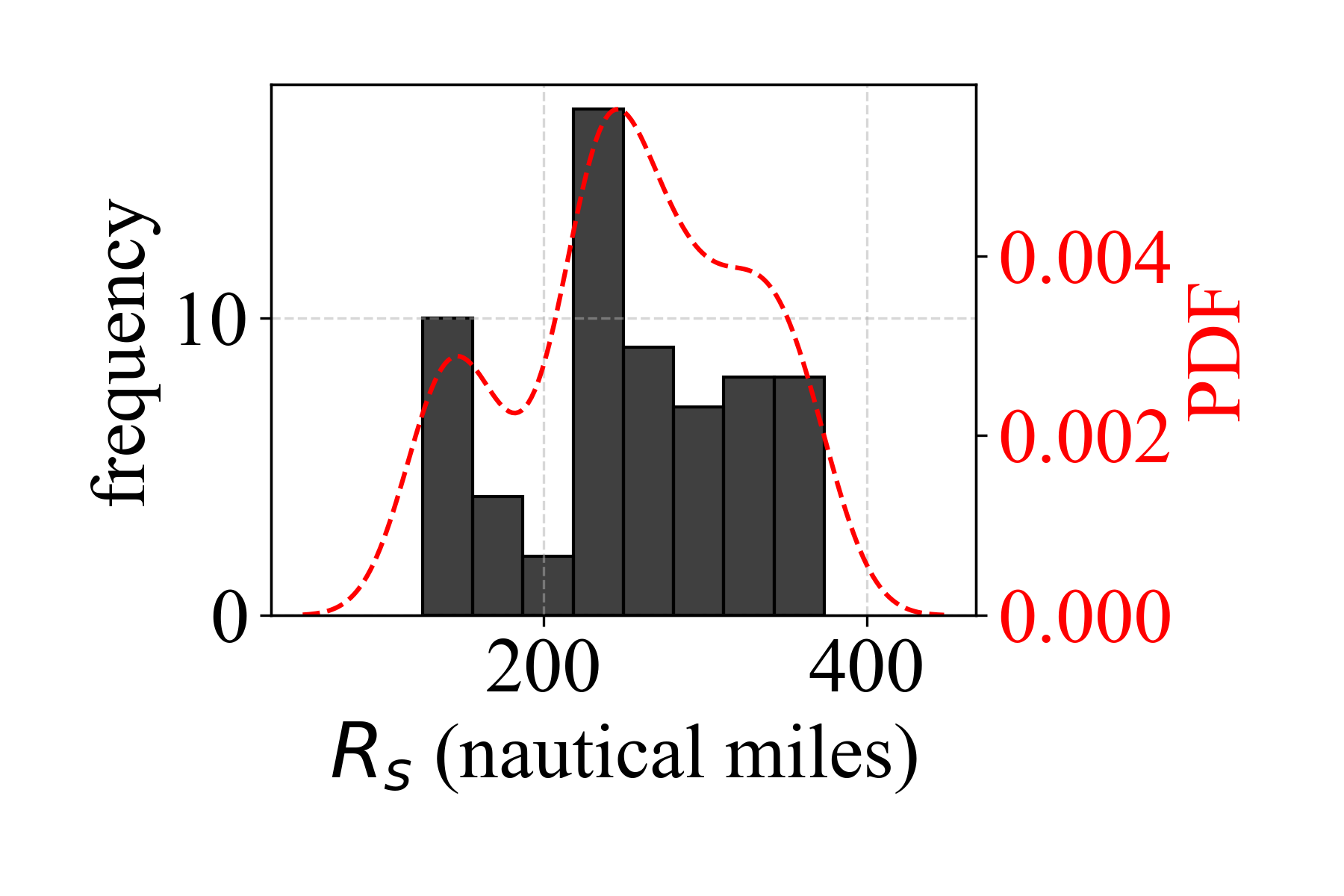}
        \vspace{-20pt}
        \caption{}
    \end{subfigure}
    \bigskip
    \hspace*{\fill}
    \begin{subfigure}[t]{0.49\linewidth}
        \centering
        \includegraphics[clip, trim=0cm 0.7cm 0.1cm 0.5cm,width=\linewidth]{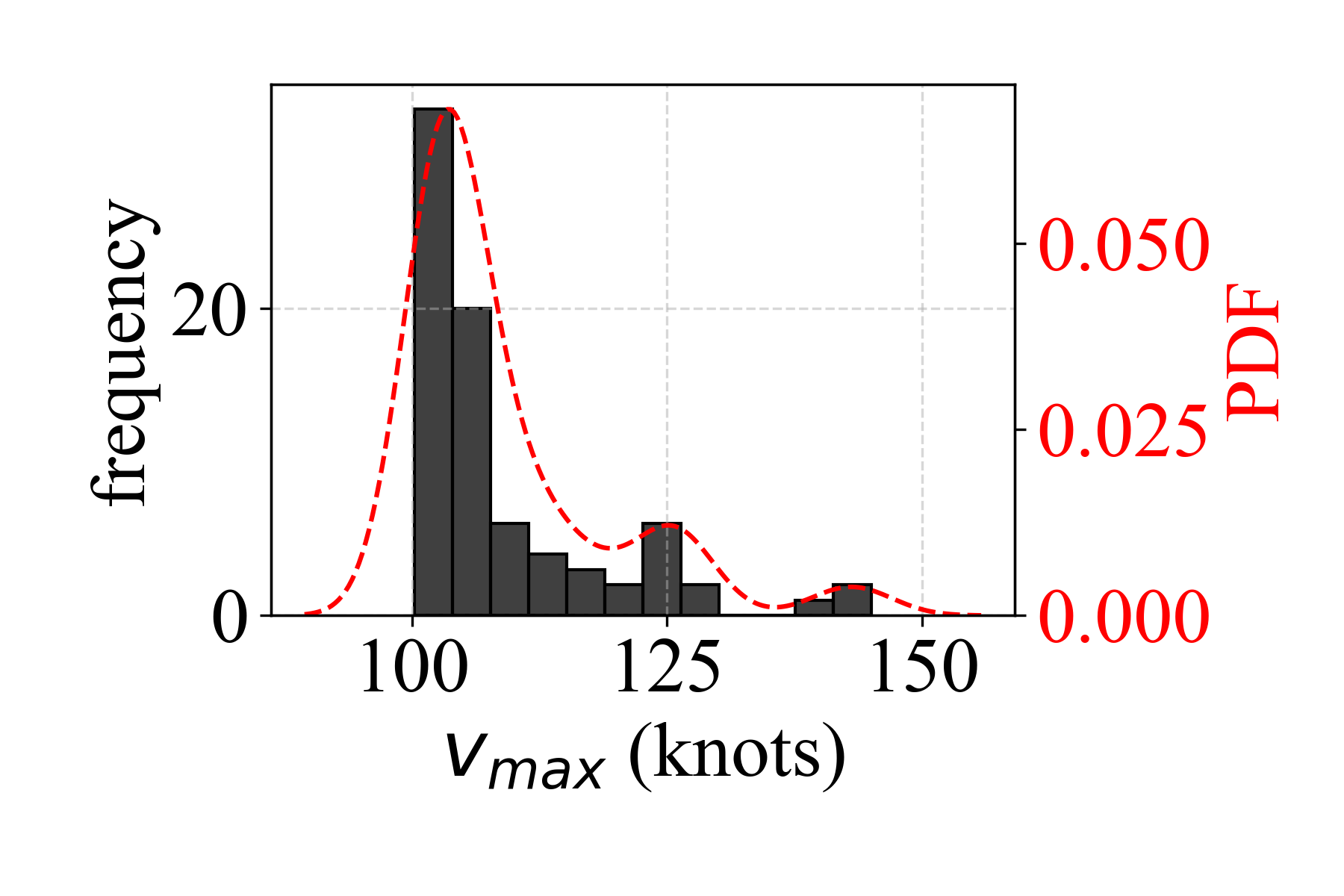}
        \vspace{-20pt}
        \caption{}
    \end{subfigure}
    \hspace*{\fill}
    \vspace{-15pt}
    \caption{KDE of a) $R_{v_{max}}$, b) $R_s$, and c) $v_{max}$ estimated from the actual hurricanes that occurred in the USA~(1851--2012).} 
    \label{fig:three graphs}
    \vspace{-15pt}
\end{figure}

The hurricane simulation is performed based on methods discussed in~\cite{Javanbakht2014, 2012}. As proposed in~\cite{Javanbakht2014}, a statistical analysis of historical data of US hurricanes (category 3 and above) from 1851 to 2012 is done and presented in Fig.~\ref{fig:three graphs}~\cite{KDE_hurricanes}. It is determined that the hurricane parameters are uncorrelated and can be randomly sampled for scenario generation purpose~\cite{Javanbakht2014}. The kernel density estimates (KDE) facilitate a method of randomly sampling the hurricane parameters. Consequently, using KDE of the hurricane parameters, $N_h=30$ equally probably hurricane scenarios are generated at the landfall location $\phi^0 = 28.9^0 N$ and $\psi^0 = 95.2^0 W$ (South East Texas on the map), which corresponds to $t = 0$ hours. In this work, we consider that the hurricanes have fixed translational speed when traversing inland. The simulations are performed every two hours for a total of twelve hour time-period, i.e., $t = [0:2:12]$ hours. It is also assumed that the hurricane forecast information is available from National Oceanic and Atmospheric Administration (NOAA)~\cite{NOAA}. Hence, the eye locations are pre-defined for each time-steps. Three equally probable tracks are considered for the hurricane, $N_\zeta = 3$. The coordinates for the eye locations of hurricanes on each of the tracks are taken from~\cite{Javanbakht2014}. All hurricanes travel inland from its landfall and decay at $t = 12$ hours for each $\zeta$.   


\begin{figure*}
    \centering
        \begin{subfigure}[t]{0.33\textwidth}
            \centering
            \includegraphics[clip, trim=0cm 0.2cm 0.2cm 0.8cm,width=\linewidth]{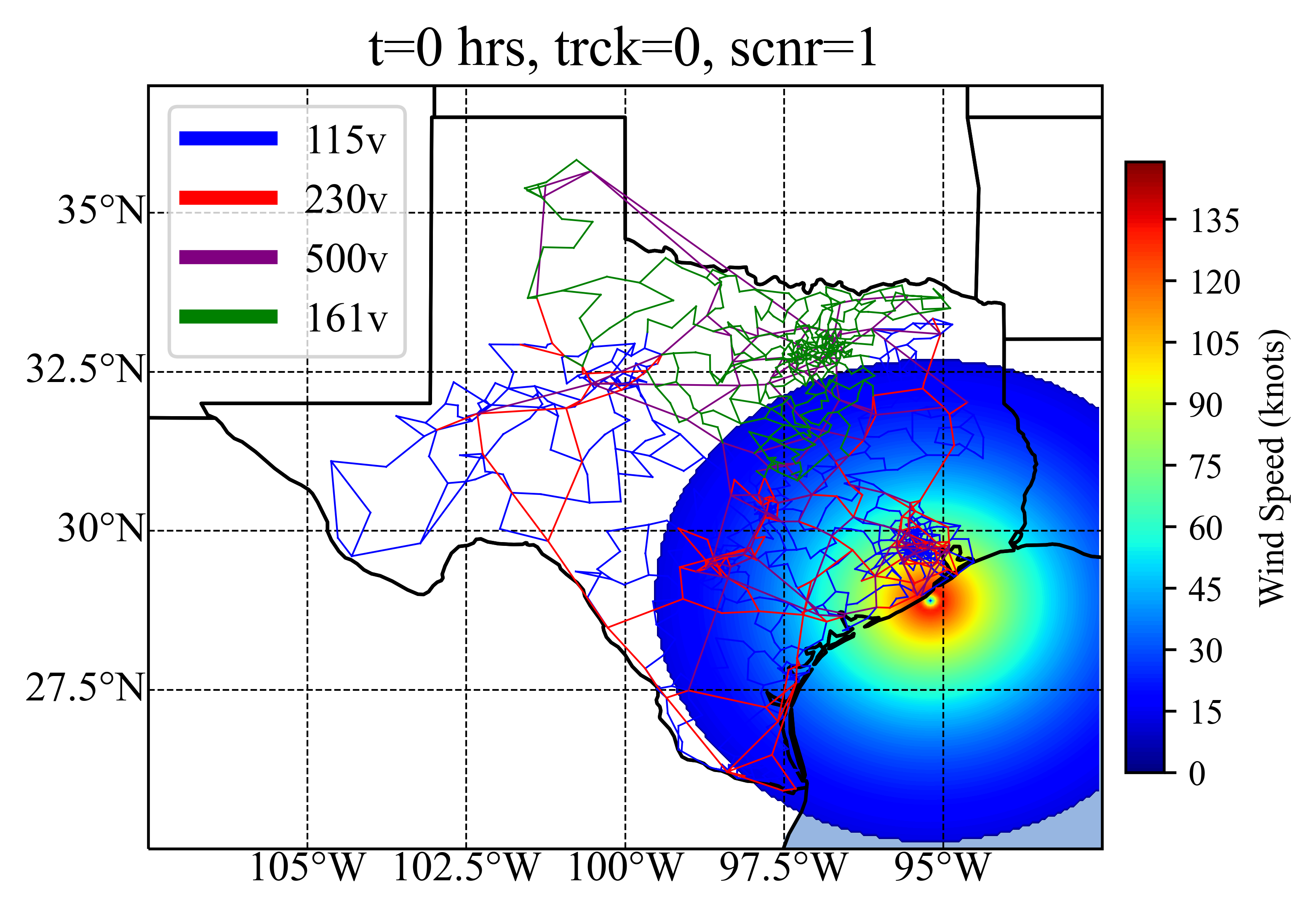}
            \caption{}
        \end{subfigure}%
        \hspace*{\fill} 
        \begin{subfigure}[t]{0.33\textwidth}
            \centering
            \includegraphics[clip, trim=0cm 0.2cm 0.2cm 0.8cm,width=\linewidth]{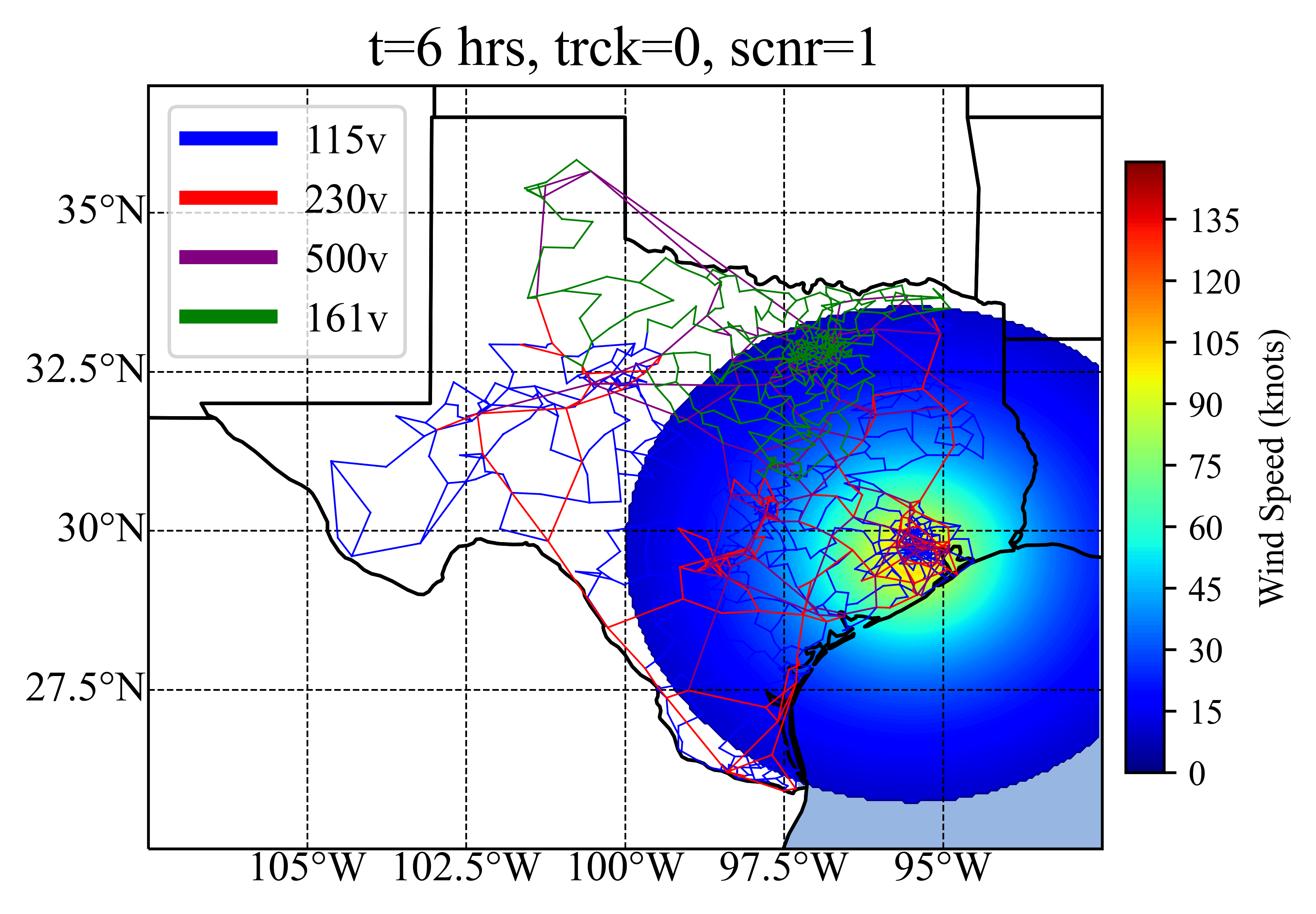}
            \caption{}
        \end{subfigure}%
        \hspace*{\fill}
        \begin{subfigure}[t]{0.33\textwidth}
            \centering
            \includegraphics[clip, trim=0cm 0.2cm 0.2cm 0.8cm,width=\linewidth]{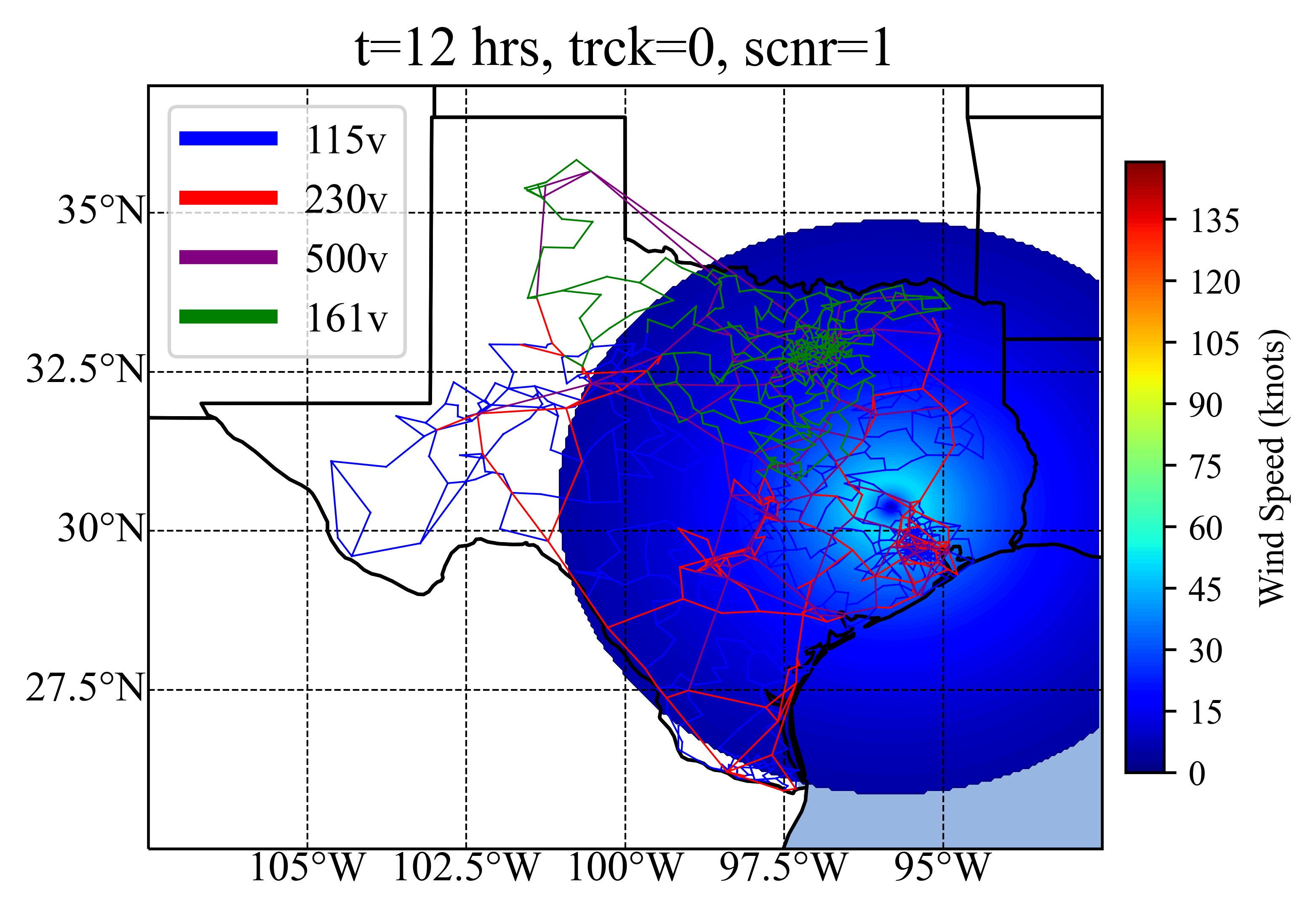}
            \caption{}
        \end{subfigure}%
        \hspace*{\fill}
    \caption{Spatiotemporal hurricane simulated on the footprint of Texas and ACTIVSg2000 test case for $\zeta = 1$ and $h = 2$. a) $t = 0$ hours b) $t = 6$ hours c) $t = 12$ hours}
    \label{fig:hurricane_model}
    \vspace{-15pt}
\end{figure*}
\subsection{Power Grid Model and Line Outage Probability}
For the power grid model, we use ACTIVSg2000: 2000-bus synthetic Texas grid developed by~\cite{7725528}. The model also contains geographical information of the buses and lines mapped on the footprint of Texas. The grid contains 1918 transmission lines (excluding internal substation connections) at 4 different voltage levels --- 115 kV, 161 kV, 230 kV, and 500 kV. Furthermore, it has 544 generators with a generating capacity of 96291.5 megawatt (MW) and 1125 loads with a total size of 67109.2 MW. Once the hurricane scenarios are generated, $d_{min}^{t,\zeta}(l,h)$ and $d_{max}^{t,\zeta}(l,h)$ are calculated. Without loss of generality, it is assumed that the minimum distance from the hurricane eye to a transmission line is the perpendicular distance and the maximum distance is the distance from the eye to one of the edges of the transmission line. Using this information and (\ref{eq:HIM}), the maximum wind speed for each transmission line is obtained as given by $\Gamma_{l, h}^{t,\zeta}$. The outage probability can then be calculated by mapping $\Gamma_{l, h}^{t,\zeta}$ with respective outage probability (\ref{eq:outage}).    

\subsection{Impact Assessment through Monte-Carlo Simulation}
MCS is performed for each $\zeta$, $h$, and $t$. For each MCS trial, $\mathbb{P}_{out}^{t,\zeta}(l,h)$ is compared with a random number $r\sim U(0,1)$. If $\mathbb{P}_{out}^{t,\zeta}(l,h) > r$, then  $\delta_{l,h}^{t,\zeta}$ is set to 1 meaning that the line is out of service. This information is sent to the power grid simulator and the corresponding line is removed from the power grid model. Here, the loss is calculated as the amount of load that is disconnected or islanded from the main grid. At the end of each trial, a moving average of the loss is calculated. The simulation is run until the moving average of the loss converges to some value. 
For the next time step, it is ensured that $\delta_{l,h}^{t+1,\zeta} \geq \delta_{l,h}^{t,\zeta}$. This means that if a line $l$ is out of service for a hurricane scenario $h$ at time step $t$ and track $\zeta$, it must be out of service for future time steps $t+1$ for the same $h$ and $\zeta$. Therefore, through MCS, we get the average loss, $loss^{t,\zeta}_{h_{avg}}$, for a given $h$ and $\zeta$ at each time step, $t$.

\subsection{Spatiotemporal Loss Quantification}
The system loss calculated using MCS is valid for different time steps $t$. However, the calculation is based on specific $\zeta$ and $h$. We assume that all $N_h$ possible hurricane scenarios and $N_\zeta$ possible tracks have an equal probability of occurrence. Hence, for a specific $t$, the potential system is calculated using (6).

\vspace{-15pt}

\begin{equation}
    loss^t = \frac{1}{N_\zeta \times N_h}\sum_{\zeta = 1}^{N_\zeta}{\sum_{h = 1} ^ {N_h} {loss_{h_{avg}}^{t,\zeta}}}
\end{equation}

\noindent
Here, $loss^t$ incorporates the spatial effect of $N_h$ hurricanes generated over $N_\zeta$ tracks. Thus, $loss^t$ is calculated for each time step as the hurricane moves inland. 

\section{Results and Analysis}\label{sec:results}
The hurricane scenarios are generated in Python and loss calculation, along with MCS, is performed in MATLAB 2021a with power grid modeled in MATPOWER 7.1~\cite{5491276}. The hurricanes and the transmission lines of ACTIVSg2000 are mapped on the footprint of Texas using a Python package designed for geospatial data processing called Cartopy~\cite{Cartopy}.    

\subsection{Simulating Hurricanes on the Power Grid}
Fig.~\ref{fig:hurricane_model} shows the hurricanes simulated on the geographical footprint of Texas along with the transmission lines for three different time steps, $t = {0, 6, 12}$ hours respectively. A particular scenario and track of hurricane, $\zeta = 1, h = 2$, is shown here. It can be seen that the wind field of the hurricane decays as it travels inland and has limited or no effect on the transmission lines once it reaches $t = 12$ hours. Furthermore, it is also to be noted that the hurricane parameters, as discussed before, are uncorrelated. Therefore, it is clear from Fig.~\ref{fig:hurricane_model} that although the size of the hurricane, $R_s$, is bigger for $t = 12$ hours, the wind field at that time step is almost decayed and has no effect on the transmission lines. 

\subsection{Impact Assessment and Loss Quantification}
\begin{figure}
    \centering
    \begin{subfigure}[t]{0.235\textwidth}
        \centering
        \includegraphics[width=\linewidth]{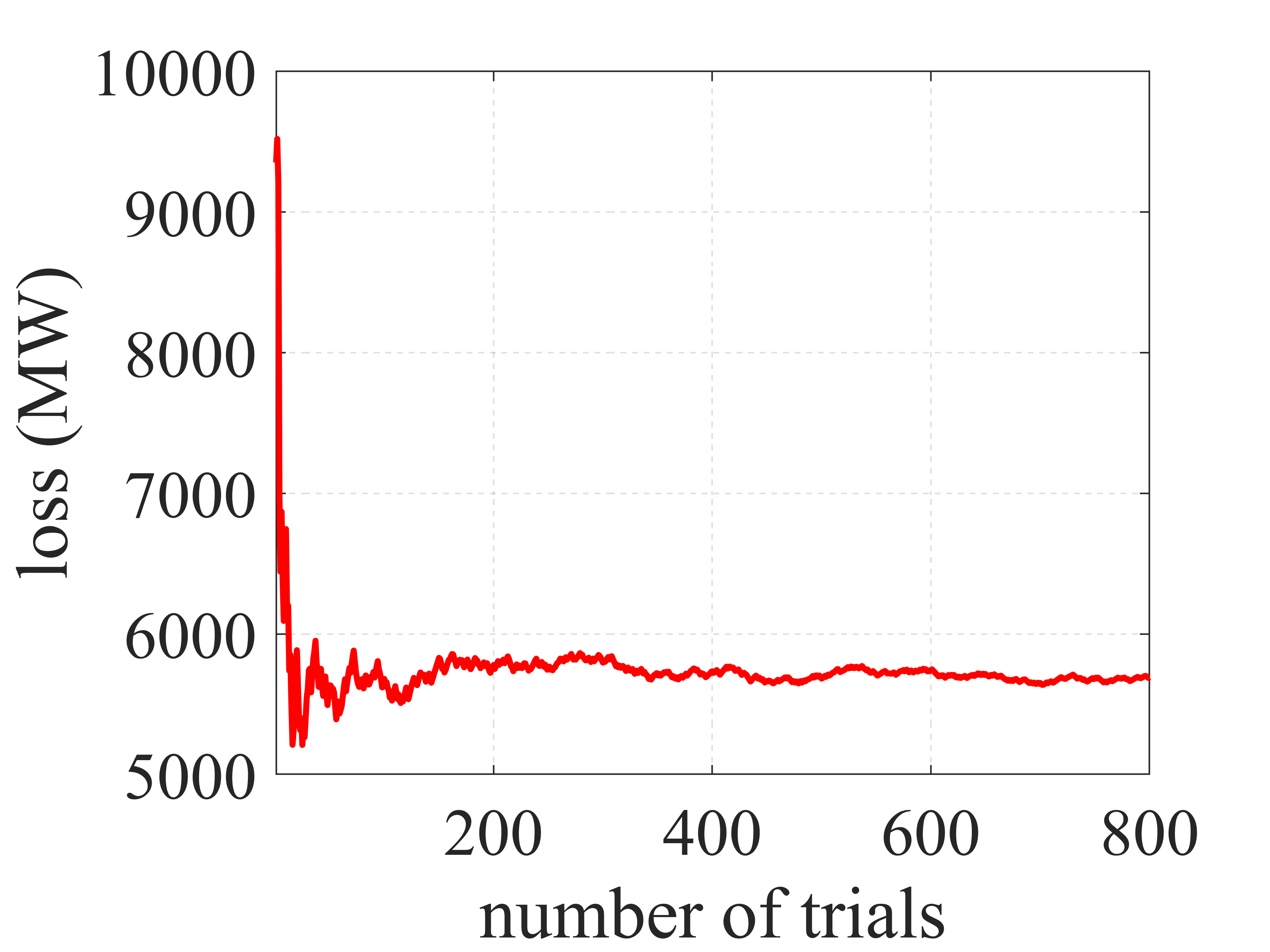}
        \caption{}
        \label{fig:MC_trial}
    \end{subfigure}%
    \hspace*{\fill}
    \begin{subfigure}[t]{0.235\textwidth}
        \centering
        \includegraphics[width=\linewidth]{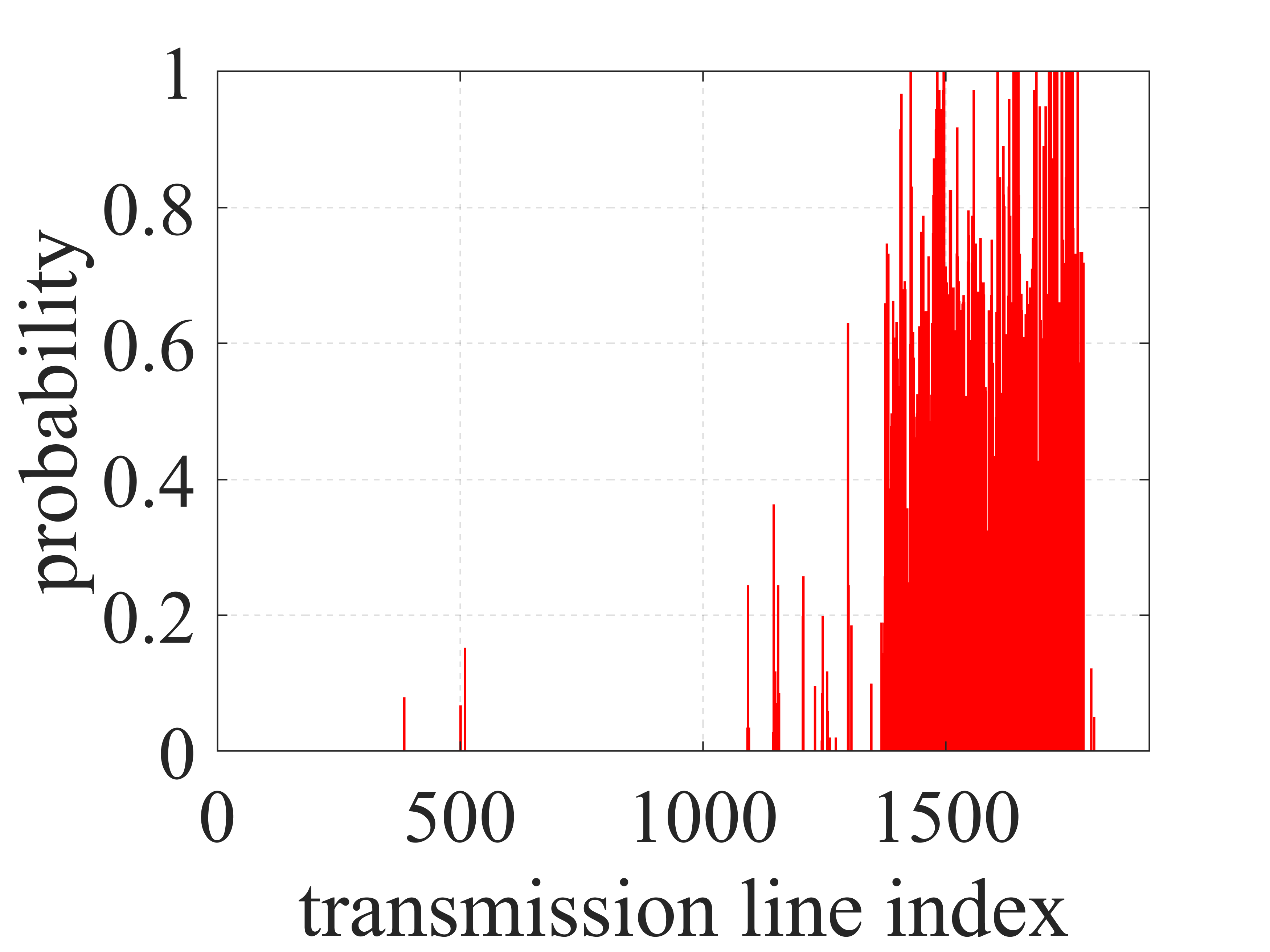}
        \caption{}
        \label{fig:outage_prob}
    \end{subfigure}
    \bigskip
    \begin{subfigure}[t]{0.235\textwidth}
        \centering
        \includegraphics[width=\linewidth]{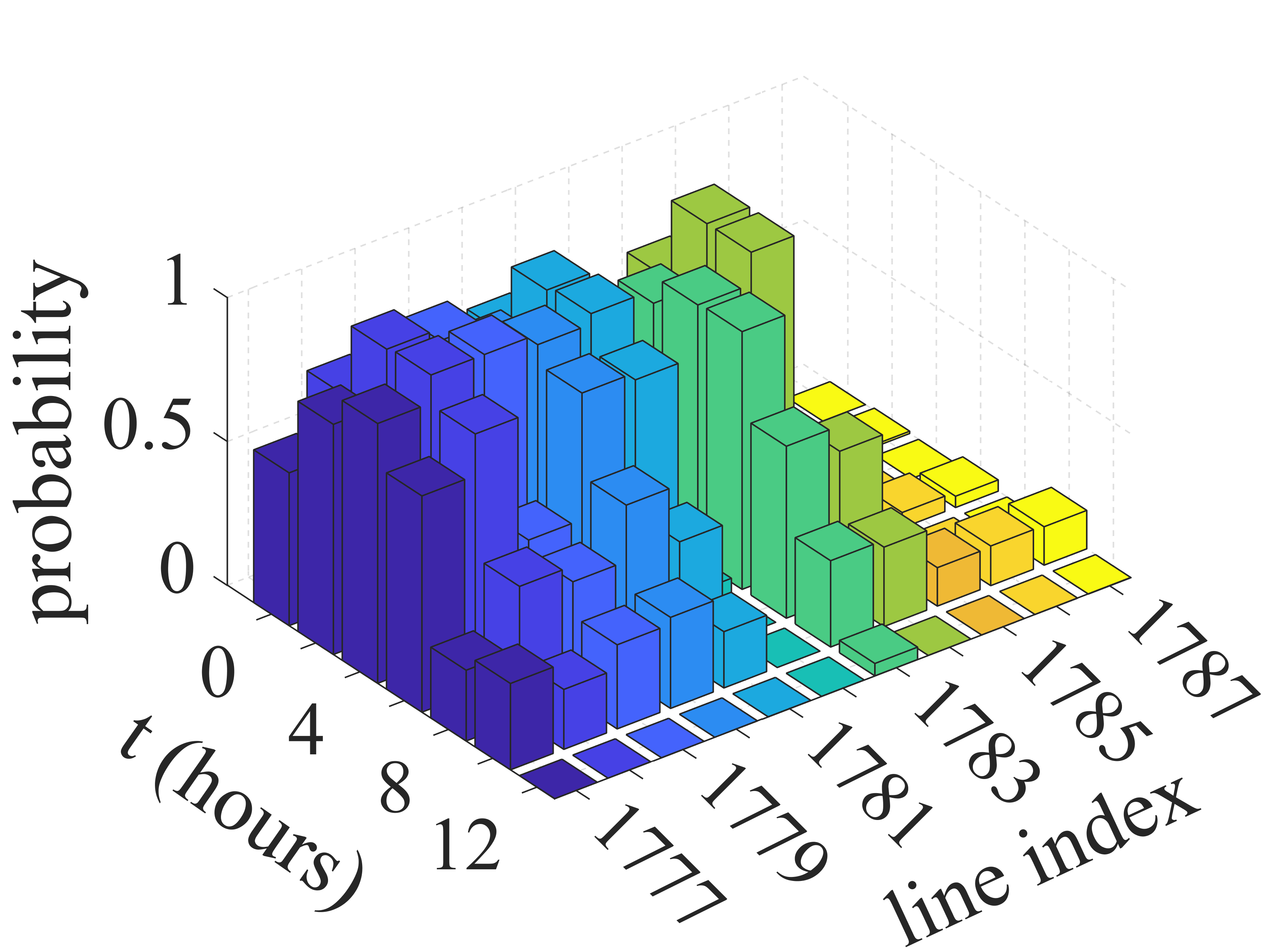}
        \caption{}
        \label{fig:3D}
    \end{subfigure}
    \hspace*{\fill}
    \begin{subfigure}[t]{0.235\textwidth}
        \centering
        \includegraphics[width=\linewidth]{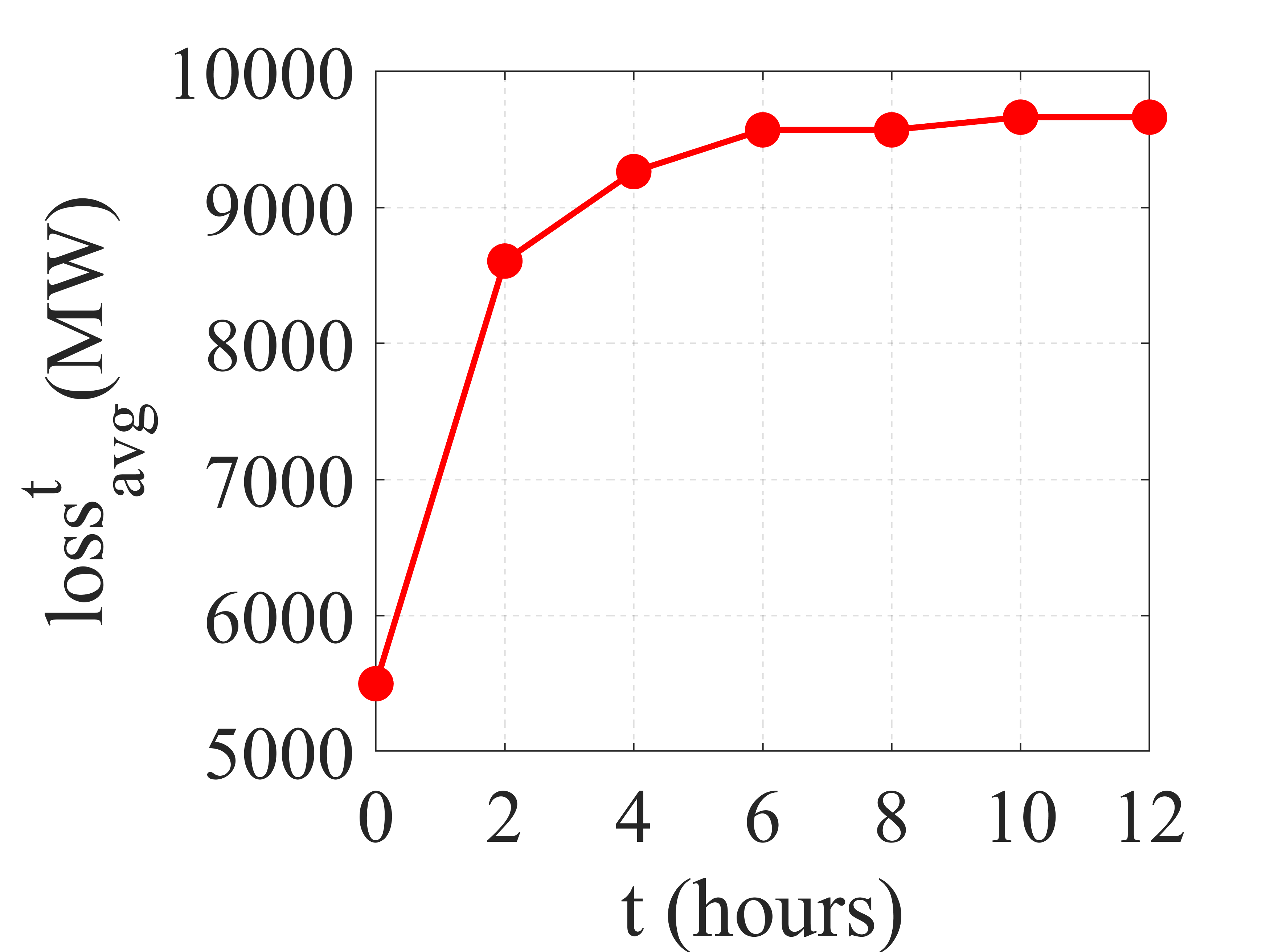}
        \caption{}
        \label{fig:loss_scen}
    \end{subfigure}
    \vspace{-10pt}
    \caption{a) MCS-based loss calculation for $t = 0$ hours. b) $\mathbb{P}^{t}_{out}$ for different transmission lines for $t = 0$ hours. c) average loss for each $t$. d) $\mathbb{P}^{t}_{out}$ for each $t$ for a few transmission lines. For all of the above cases, $\zeta = 1$, and $h = 2$.} 
    \vspace{-15pt}
\end{figure}

Fig.~\ref{fig:MC_trial} shows MCS of 800 trials for specific outage probabilities of transmission lines obtained when $\zeta = 1$, $h=2$, and $t=0$ hours. Through experimentation, it was concluded that 800 trials are enough for convergence of the loss for any simulation case. For this case, $loss^0_{avg} = 5498.13$ MW. Fig.~\ref{fig:outage_prob} shows the outage probability values, $\mathbb{P}^{t}_{out}$, for which the MCS shown in Fig.~\ref{fig:MC_trial} is performed. The transmission lines which are near the location of the landfall have higher $\mathbb{P}^{t}_{out}$. This is due to the fact that the hurricanes have high intensity wind field contours around the location of the landfall. The value of $\mathbb{P}^{t}_{out}$ decreases, as shown in Fig.~\ref{fig:3D}, as the intensity of the hurricane decreases while moving inland. Furthermore, Fig.~\ref{fig:loss_scen} presents the average loss obtained from MCS, $loss^t_{avg}$, for each $t$, $\zeta = 1$, and $h=2$. The loss at $t = 0$ hours corresponds to the converged value of loss from Fig.~\ref{fig:MC_trial}. The loss increases as the hurricane moves forward, $loss^2_{avg} = 8605.73$ MW, $loss^6_{avg} = 9570.44 $ MW, and finally settles at $loss^{12}_{avg} = 9663.84$ MW.          

\begin{figure}[ht]
    \centering
    \includegraphics[width=0.8\linewidth]{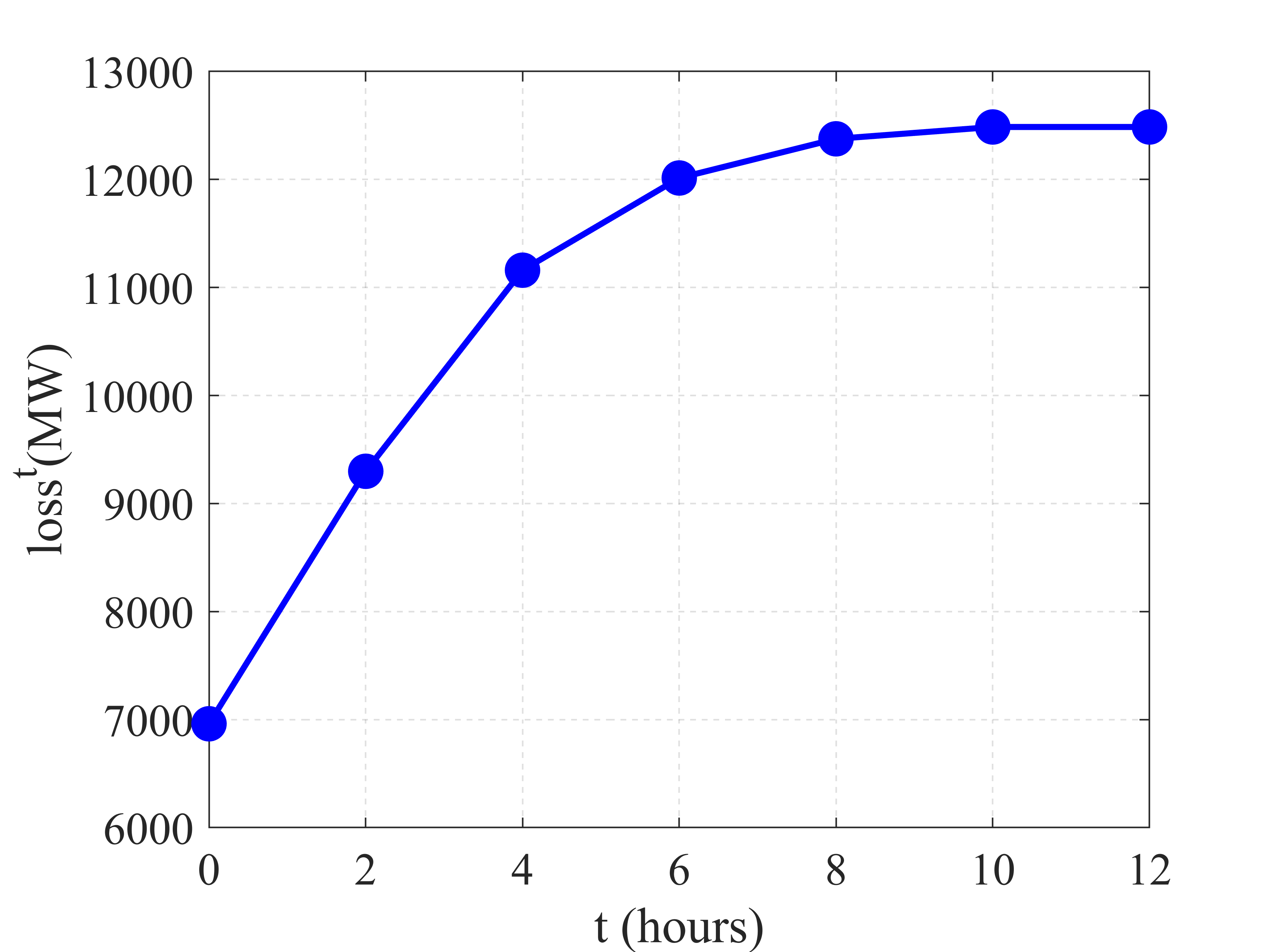}
    \caption{Time-varying impacts of stochastic hurricane scenarios.}
    \label{fig:overall_time_loss}
    \vspace{-15pt}
\end{figure}

Fig.~\ref{fig:overall_time_loss} shows the overall loss for each $t$ calculated for $N_h$ hurricanes and $N_\zeta$ tracks. Consistent with the single hurricane simulated at single track scenario, $loss^{t}$ increases from $t=0$ hours, $loss^{0}=6961.05$ MW, to $t=6$ hours, $loss^{6}=12013.60$ MW, and slowly saturates as it reaches around $t=12$ hours, $loss^{12}=12485.86$ MW. As discussed above, the initial ramp in the loss is due to high intensity wind field contours of hurricanes around the location of landfall and decays gradually as it moves inland.   

\section{Conclusion}\label{sec:conclusion}
In this paper, the spatiotemporal impacts of hurricanes on the power grid are analyzed and quantified based on the total loads disconnected from the grid. The simulation results showed that the most prominent effects of hurricanes were observed during the time of their landfall, and the effects sustained for a few more time steps as they moved inland. Nonetheless, hurricanes decay after some time and have a limited impact on the power grid. The system loss increases as time progress, and the hurricanes damage several lines as traversing forward. The proposed spatiotemporal loss quantification metric helps better understand the possible impacts of hurricanes on the power grid at a finer space and time granularity.  The metric can be used to develop proactive planning strategies to mitigate and reduce the impacts of hurricanes considering the time and space-dependent losses for the power systems. 



\bibliographystyle{IEEEtran}
\bibliography{references} 
\end{document}